\documentclass[aps,floats,superscriptaddress,showpacs]{revtex4}
\usepackage{amsmath}
\usepackage{amsfonts}
\usepackage{amssymb}
\usepackage{color}

\usepackage{epsfig}
\usepackage{graphicx}
\usepackage{dcolumn}
\usepackage{bm}

\newcommand{\beq}{\begin{equation}}
\newcommand{\eeq}{\end{equation}}
\newcommand{\beqa}{\begin{eqnarray}}
\newcommand{\eeqa}{\end{eqnarray}}

\def\gapp{\lower.35em\hbox{$\stackrel{\textstyle>}{\sim}$}}
\def\lapp{\lower.35em\hbox{$\stackrel{\textstyle<}{\sim}$}}

\begin{document}
\bibliographystyle{apsrev}

\title{Aharonov-Bohm interferences from local deformations in graphene}
\author{F. de Juan}
\affiliation{Department of Physics, Indiana University, Bloomington, IN 47405}
\author{A. Cortijo}
\affiliation{Departamento de F\'isica Te\'orica, Universidad Aut\'onoma de Madrid, E-28049, Madrid,
Spain}
\author{M. A. H. Vozmediano}
\affiliation{Instituto de Ciencia de Materiales de Madrid (CSIC),
Cantoblanco, Madrid 28049, Spain}
\author{A. Cano}
\affiliation{European Synchrotron Radiation Facility, 6 rue Jules Horowitz, BP 220, 
38043 Grenoble, France}

\date{\today}
\begin{abstract}
One of the most interesting aspects of graphene is the tied relation between structural and electronic properties.  The observation of ripples in the graphene samples both free standing and on a substrate 
has given rise to a very active investigation around the 
membrane-like properties of graphene and the origin of the ripples remains as one of the most interesting open problems in the system. The interplay of structural and electronic properties is successfully described by the modelling of curvature and elastic deformations by fictitious gauge fields 
that have become an experimental reality after the suggestion  
that  Landau levels can form associated to strain in graphene  
and the subsequent experimental confirmation.   
Here we propose a  device   to detect microstresses in graphene based on a  scanning-tunneling-microscopy setup able to measure Aharonov-Bohm interferences at the nanometer scale. The interferences to be  observed in  the local density of states are created by the fictitious magnetic field associated to elastic deformations of the sample.

\end{abstract}
%
%

\maketitle

\section{Lattice deformations and fictitious magnetic fields in graphene.}

Since graphite monolayers started to be isolated in a controlled way \cite{Netal05,Zetal05}, graphene has been an optimal playground to test the most exciting  ideas in condensed matter  \cite{RMP08}. A great deal of attention was initially paid to its striking electronic properties, but it was soon realised that the structural and mechanical   properties can be even more interesting both from a fundamental point of view as well as for applications \cite{GN07}. The modelling of curvature  by gauge fields in graphene was suggested in the early publications associated to topological defects needed to form the fullerene structures \cite{GGV92,GGV93}. The main idea was that the phase acquired by an electron circling a pentagonal defect is the same as that arising when circling a solenoid with the appropriate magnetic flux in analogy with the Aharonov--Bohm effect. The fictitious magnetic fields were later applied to model the observed ripples \cite{Metal07,Metal07b,Setal07}  and elastic deformations \cite{MNetal06,MG06,JCV07,GHL08,KC08,GKV08}.  The state of the art and an updated list of references can be found in  \cite{VKG10}. 

A turn of screw took place when the fictitious magnetic field became an experimental reality after the suggestion   that Landau levels can form associated to strain in graphene  \cite{GKG10,GGetal10} and their subsequent observation in \cite{LBetal10}. The formation of Landau levels requires  very high values of the  fictitious magnetic fields and hence very strong deformations  of the samples. Fields of up to 300 Tesla were estimated for the observed  nano bubbles in \cite{LBetal10}. In the present work, we consider the opposite limit where the fields are small and the electronic excitations can still be described in terms of plane waves (rather than Landau levels). As we show below, non-trivial amusing effects can also be generated by fictitious gauge fields in this limit. Specifically, we discuss the realisation of the Aharonov-Bohm (AB) effect via deformation fields and propose a simple scanning-tunneling-microscopy (STM) setup that can make use of this realisation to detect stresses. 

\section{Aharonov-Bohm interferences from fictitious gauge fields.}
\label{ABI}
\begin{figure}
\begin{center}
\includegraphics[height=8cm]{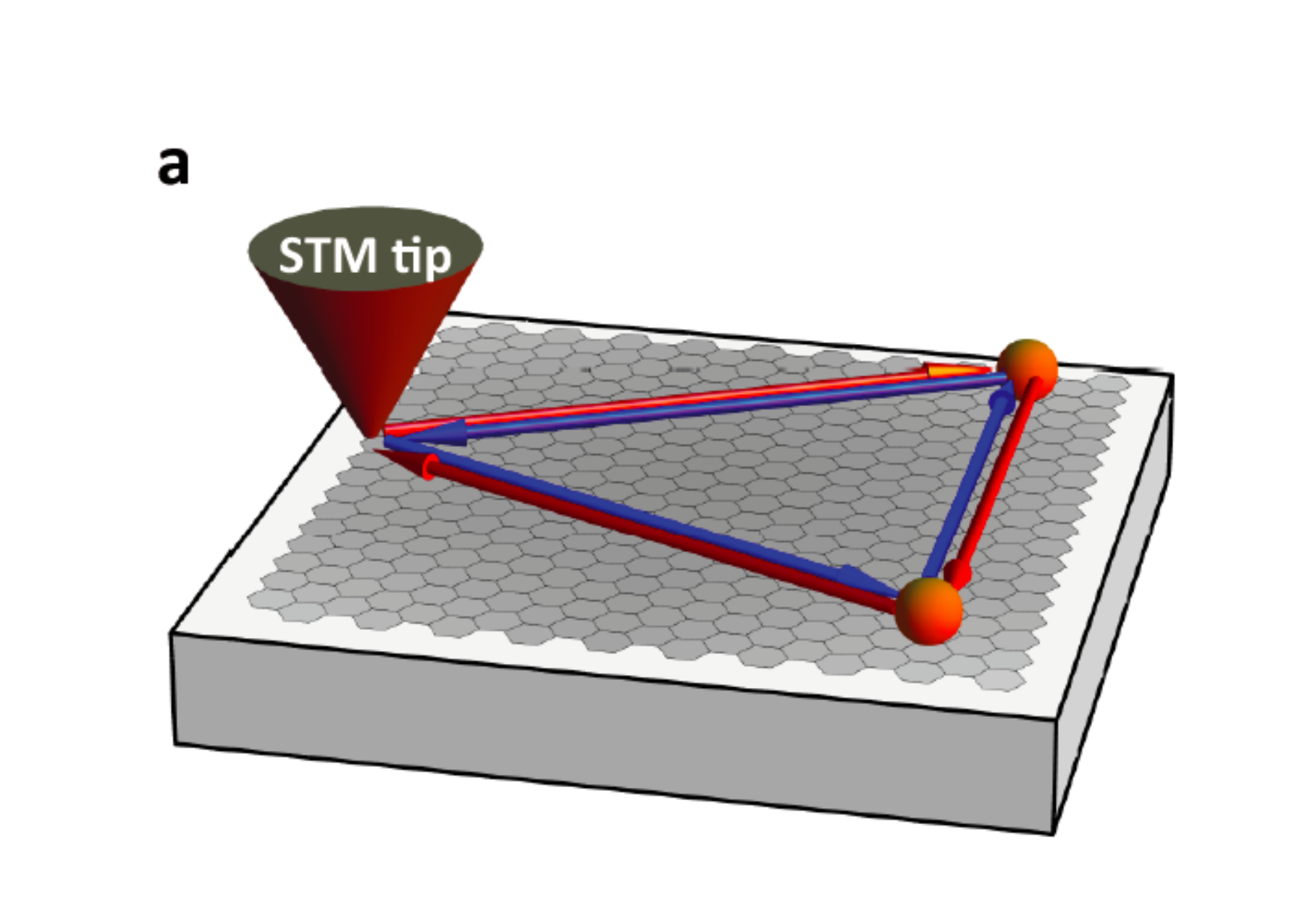}
\hspace{0.5cm}
\includegraphics[height=5cm]{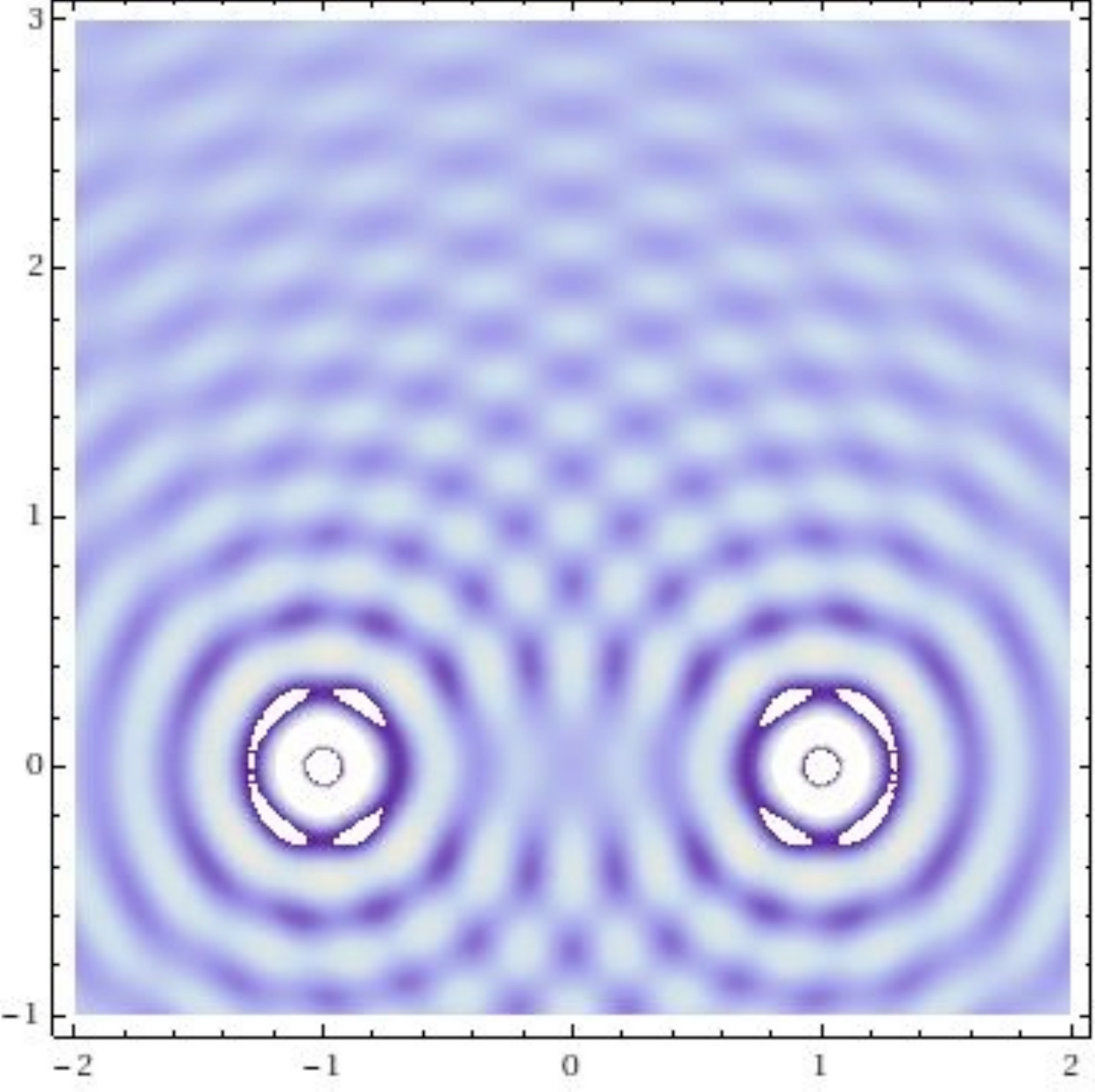}
\caption{\label{Fig1a} Basic sketch of the proposed device. a. The two scatterers (red balls)
and the STM tip design a semiclassical closed path that electrons can follow
clockwise or anticlockwise (red and blue lines in the figure). If the region
enclosed by the path is strained or curved, the fictitious magnetic field
associated to the deformation will give rise to a characteristic interference
pattern in the LDOS arising from the asymmetry of the clockwise and anticlockwise
paths induced by the fictitious field as described in the text (see also fig. 2). b Typical LDOS map
to be observed in the STM experiment when the region enclosed is perfectly
flat.}
\end{center}
\end{figure}

The AB effect \cite{AB59} is one of the hallmarks of quantum  physics and has continued awaking  interest along the years \cite{WW86,ODetal98}, partly due to the difficulty of finding appropriate experimental realizations. In our setup, the fictitious gauge fields  arise from strained (or curved) portions of the graphene sheet and the AB interference manifests in the form of oscillations in the local density of states which can be detected by a STM. A simple way to generate this interference is by introducing scattering centres and thus engineering the scattering loop depicted in Fig. \ref{Fig1a}a.  The electrons injected by the STM tip can follow this loop either clockwise or anticlockwise, which results in different phases due to the fictitious fields inside the loop. This setup does not require the two physical paths of conventional AB devices, and is analogous to the one recently proposed in \cite{CP09} to detect nanoscale variations of  magnetic fluxes. In what follows we will analyse the different aspects of the fictitious-field case and show the experimental feasibility of our proposal. 

Time-reversal symmetry breaking is a key point in the magnetic AB effect. This gives rise to the additional AB phase in the electron wave function that, for closed trajectories, is proportional to the magnetic flux traversing the area enclosed by the loop. Thus, closed trajectories travelled clockwise and anticlockwise produce phases with opposite sign, which can further be exploited as described in \cite{CP09}, for example, to build a STM AB interferometer. In the case of elastically deformed or curved graphene, however, at first glance it seems that the fictitious fields will give no net AB effect because these deformations do not break time-reversal symmetry. The symmetry is preserved by the fictitious magnetic fields because these fields couple with opposite signs to the two different Dirac cones of graphene \cite{GGV93} (see the supporting  information). In the following we show that, contrary to this initial expectation, a non-zero AB phase does arise from the elastic deformation fields in graphene. 

To our purposes it is convenient to write the electron wave function as  $\Psi^+  = (\Psi_{K}, \Psi_{K'})$ where $\Psi_{K(K')}$ is the bi-spinor corresponding the Fermi point $K (K')$.  As discussed in the supplementary information, a deformation of the graphene lattice gives rise to  a fictitious gauge field 
\beq
{\bf A}= {\kappa \Phi_0 \over \pi }\begin{pmatrix} 
u_{xx}-u_{yy} \\
-2u_{xy}
\end{pmatrix}
\label{A}
\eeq
where $\kappa \simeq 3$ nm$^{-1}$, $\Phi_0$ is the flux quantum and $u_{ij}$ is the strain tensor which can be written as $u_{ij} = \frac{1}{2}[\partial_j u_i+\partial_i u_j+(\partial_i h)(\partial_j h)]$ in terms of the in-plane and out-of-plane displacements $\mathbf u$ and $h$ respectively \cite{VKG10}. The graphene wavefunction  then accumulates a phase factor that can be written as $e^{-i \pi {\Phi\over \Phi_0} \tau_3}$ for the electrons travelling along a closed path $C$, where $\tau_3 = \text{diag} (1,-1)$ is the third Pauli matrix and $\Phi =\oint_{C} {\bf A}({\bf l})\cdot d{\bf l}$ is the flux of the fictitious field that passes through the corresponding area. Consider now the closed scattering path shown in Fig. \ref{Fig1a}a and let $a_-$ and $a_+$ be the scattering amplitudes for electrons in the state $\Psi_0 $ that travel this path clockwise and anticlockwise respectively. These amplitudes can be expressed as $a_{\pm} = e^{\pm i \pi {\Phi\over \Phi_0} \tau_3} a_0$, where $a_0$ is the scattering amplitude in the absence of field. The scattered wavefunction $\Psi_s$ then will be such that 
$\langle \Psi_0 | \Psi_s\rangle = 2 a_0  \cos \big(\pi \Phi / \Phi_0 \big) $. 
In plain words, the key issue is that the opposite gauge potential couples to different types of particles (electrons at each valley) and the net effect of the opposite trajectories adds up which is the same reason that makes it possible the formation of Landau levels discussed in \cite{GKG10,GGetal10}. This physics is also very similar to the one discussed in \cite{MNetal06,MG06} in relation with the absence of weak localisation in graphene. 
This mismatch between the electron wave functions will give rise to a distinctive signal in the local density of states (LDOS) that can be measured by an STM tip as follows. 

The LDOS can be expressed as  $N({\bf r}, \omega)=-\frac{2}{\pi}\text{ Im Tr }G({\bf r}, {\bf r};\omega)$ in terms of the retarded Green's function $G$. In the presence of a (real or fictitious)  gauge  field $\mathbf A$, it follows from the previous discussion that the electron Green's function for a homogeneous graphene sheet factorizes as
\beq
G_0({\bf r}-{\bf r'}, \omega)=\exp\left( i\frac{\pi}{\Phi_0}\int_{\mathbf r}^{\mathbf r'}{\mathbf A}({\bf l})\cdot d {\bf l}\;  \right)
G_{00}({\bf r}-{\bf r'}, \omega)
\label{G0A}\eeq
in the semiclassical approximation, where $G_{00}({\bf r}-{\bf r'})$ is the Green's function for zero gauge field and the line integral goes along the straight line connecting $\mathbf{r}$ and $\mathbf{r'}$. When the graphene sheet contains two scattering centres as depicted in Fig. \ref{Fig1a}a the corresponding the LDOS turns out to be
\beq
N({\bf r}, \omega)=
N_{A=0}
({\bf r}, \omega) + N_{{\rm loop}}({\bf r}, \omega)
\left\{\cos[\pi \Phi({\bf r})/\Phi_0]-1\right\},
\label{oscdos}
\eeq
where $N_{A=0}$ represents the total LDOS in the absence of deformations, $N_{\rm loop}$ the interference of all the scattering paths passing through the two impurities and enclosing a finite area in the absence of deformations, and $\Phi$ the flux of the pseudo-magnetic field induced by the deformations through this area.  Here the $\cos $ appears due to the AB interference associated with the deformation of the graphene sheet. There is a correspondence between this interference and the one described in \cite{CP09} for the real magnetic case and non-relativistic electron systems. This correspondence, however, is rather non-trivial due to the matrix structure of the electron Green's function in graphene.  The details of the calculation that gives rise to the expression \eqref{oscdos} in graphene are given in the supplementary information.
\begin{figure}
\begin{center}
\includegraphics[height=8cm]{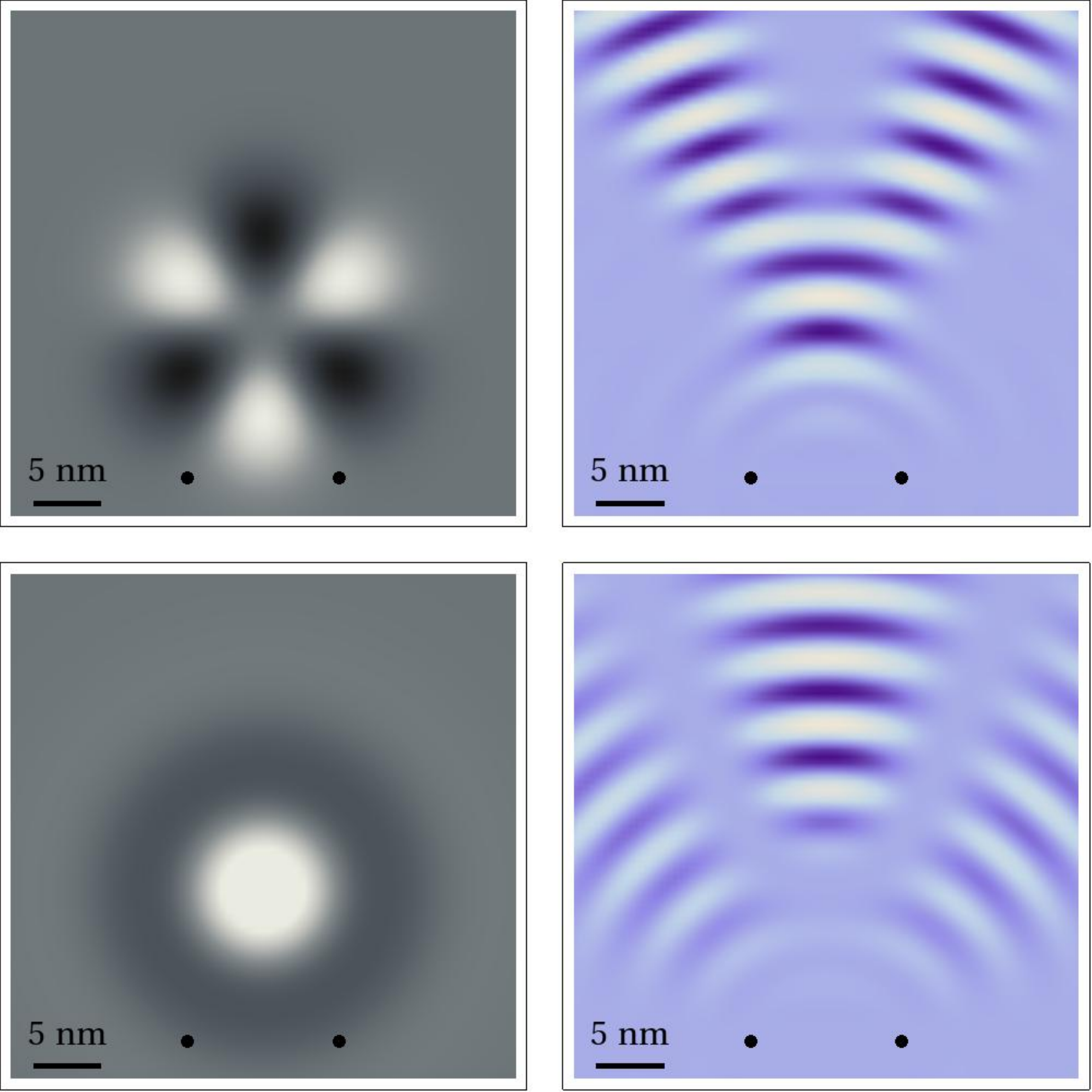}
\caption{Strain-induced pseudo magnetic fields and AB interferences in the LDOS. a. Distribution of pseudo magnetic field generated by the in-plane strain of equation (4) discussed in the text. b. Interference correction to the LDOS at energy $\omega$ = 0.4 eV after subtraction of the $N_{u=0}$ contribution, as given by eq. (3). c and d represent the same for the case of the out-of-plane gaussian height profile of equation (5) discussed in the text.}
\label{Fig2}
\end{center}
\end{figure}

\section{Experimental proposals.}

The fact that there is a real AB effect in the LDOS due to fictitious gauge fields that appear in deformed graphene is illustrated in Fig. \ref{Fig2}. In a) we consider that the deformation of the graphene sheet corresponds to in-plane displacements (in polar coordinates): 
\beq
u_r = u_0 r^2 \sin 3 \theta ,\qquad u_\theta = u_0 r^2 \cos 3 \theta 
\label{us}
\eeq
as these described in \cite{LBetal10} but localised in a region of width $\sigma =6 $ nm [$u_0 = u_{00}/\sigma^2 \exp (-r^2/2\sigma^2)$ with $u_{00}$ a constant].  

The fictitious magnetic field configuration corresponding to this situation is shown in  Fig. 2 a.  Fig. 2 b. shows the interference pattern obtained at an energy of $\omega = 0.4$ eV and $u_{00} = 0.1$ nm after subtraction of the $N_{u=0}$ contribution for the given geometry. Figs. 2 c. and d. show the same computation  for  a pure out--of--plane deformation of the graphene sheet corresponding  to a gaussian bump of profile \cite{JCV07,GTetal10}
\beq
h(r)=A\exp(-r^{2}/2\sigma^{2}),
\label{gaussian}
\eeq
We have done the calculation with $\sigma=6$ nm, $A= 0.5$ nm. These two cases illustrate rather generically what can be expected.
\begin{figure}
\begin{center}
\includegraphics[height=6cm]{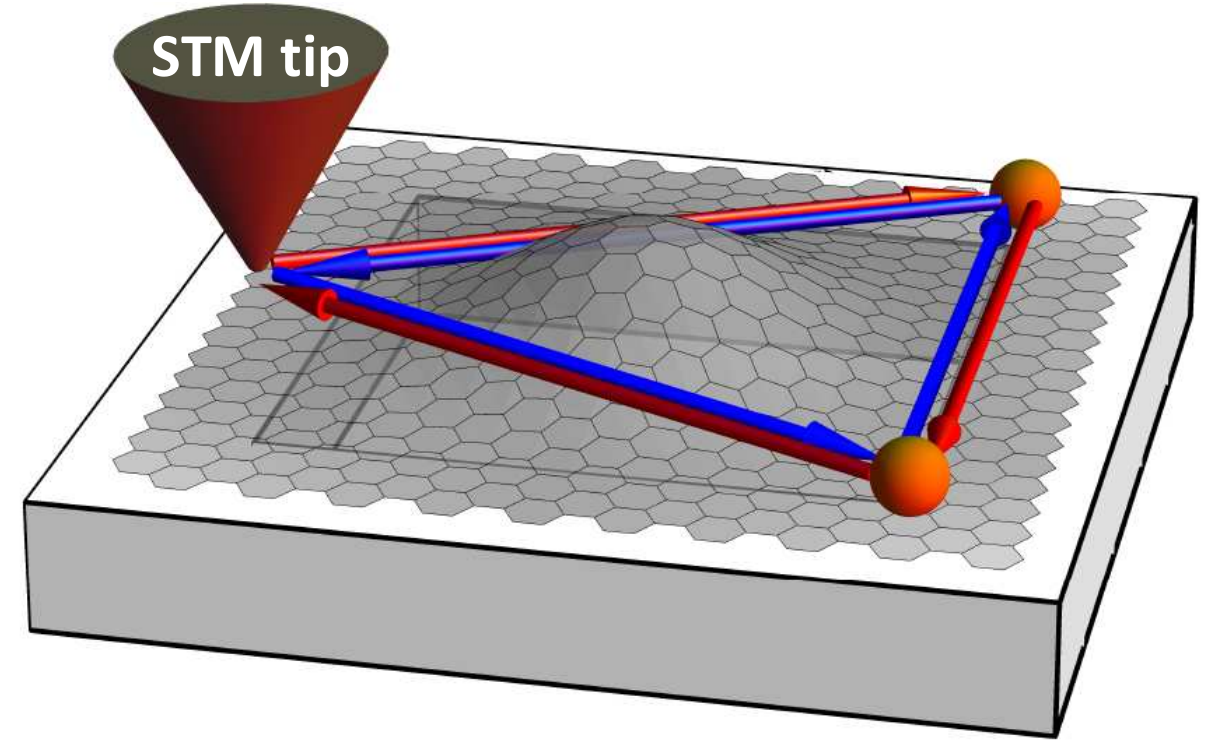}
\caption{Proposed STM interferometer as can be used to measure the strain in suspended graphene. In the figure the two impurities (red balls) are located in the sample at opposite extreme from the tip. The tip can stay at a fixed position while the fictitious magnetic field is changed by applying pressure either mechanically or with an electric field as described in \cite{FGK08}. The variation of the LDOS measured can be directly related to the mechanical deformation of the sample. The device can also be mounted on a graphene-sealed micro chamber with a small hole. When gas is  pumped inside the chamber the graphene covering the hole will inflate as a balloon as described in
\cite{BVetal08}. The STM oscillations will be a direct measure of the tension generated by pressure 
in the graphene sample.}
\label{Fig3}
\end{center}
\end{figure}

The physics discussed above opens the door to build STM devices to probe directly local stresses, and more generally mechanical properties, alternative to more indirect procedures as those using topographic measurements \cite{Tetal09} or Raman scattering \cite{MLetal09}.  It can be of use in a variety of samples to characterise strain engineering designs coming from the substrates or growth conditions \cite{SGetal07}. This can be illustrated with two point scatterers as before. A reference LDOS can be acquired by scanning with the STM tip and calibrate the device before producing the stresses in the material either thermally  \cite{BMetal09}, mechanically \cite{NWetal08}, or from the corresponding substrate \cite{SJL09}. Then, by moving the STM tip along a line parallel to the scatterers, in such a way that the area of the scattering path is conserved, a map of the corresponding deformations can be obtained by subtracting the reference LDOS. Alternatively the setup can be used to characterise the samples by subtracting the $N_{{\rm loop}}$ in Eq. (\ref{oscdos}) computed for an ideal clean, flat sample with the given positions of the scatterers. 

The suspended graphene flakes discussed in \cite{FGK08} represent a system particularly suited for exploiting the AB effect due to fictitious gauge fields (see Fig. \ref{Fig3}). These samples are under tension and the fictitious magnetic field is maximal at the edges. This field has been estimated to be of a few Tesla in sub-micron wide flakes \cite{PSetal10}. The STM tip and the scattering centres can be set in the supported (flat) region at opposite extremes from the suspended region whose mechanical deformation can be modulated by the electric force between the flake and the gate or by applying mechanical pressure. If the STM tip is let fixed at a given position, the $\cos \Phi$ oscillations of the LDOS signal gives a direct measure of the varying strain. 

Alternatively  the STM setup can be mounted on the graphene-sealed micro chamber described in \cite{BVetal08}.  There, a bump in the graphene sheet can be created in a controlled way by pumping gas in the chamber, which will generate a fictitious magnetic field localised in the region inside the bump.  Scattering centres then can be engineered in such a way that the fictitious field associated with the bump and the stresses generated here can be conveniently extracted from the LDOS. In particular, by placing the scatterers in the flat region away from the bump and performing STM scans parallel to them the area of the scattering loop is kept constant.  Thus, a first scan done before creating the bump will show typical Friedel oscillations due to the impurities. Then, when a second scan is done along the same line after the gas in pumped and the bump forms, a different interference pattern will be observed in the $dI/dV$ curves from which the distribution of pseudo-magnetic flux can be easily extracted. 

We note that, since the AB effect is due to the phase of the wave function, other candidates to produce AB interferences in graphene are topological defects generated by substitution of a hexagonal carbon ring by a different polygon.  These defects are very frequent in carbon nanotubes and other graphitic structures, pentagons and heptagons being energetically the most favourable. The fictitious fields created by these type of defects are non-abelian in the valley and sub--lattice spaces since they break the sub--lattice symmetry as described in  \cite{GGV93}. In this context, non abelian means that the vector potential has an additional matrix structure acting in the valley space. The corrections to the density of states associated with these topological defects were analysed within the covariant formalism in \cite{CV07b}. The analysis of section 2 can be easily extended to deal with these fields. Taking into account that the square of extra matrix associated to the valleys (also a Pauli matrix) is the identity, our STM setup can also be used to probe the topological defects mentioned above. The phase interference will be proportional to the deficit angle of the defect and this possibility was already discussed in \cite{LC04}. However, a possible drawback in the case of topological defects is that the effective magnetic field being strong and localised -- they are in fact like solenoids -- may act as an extra scattering centre spoiling the simplicity of the analysis performed in the present work.

\section{Physical feasibility of the proposal.}
The physical feasibility of the proposed experiments relies on the capability to select and manipulate the appropriate scatterers and, most importantly, to ensure the validity of the semiclassical approximation. In the following we discuss this feasibility. 

Different sorts of scatterers can be considered for the setup.  Vacancies, adsorbed atoms or even two passive STM tips can be used nowadays in a controlled way \cite{CL10,YTetal10,LTB10}. Inter--valley scattering will not suppress the effect but  it should be minimised for the best performance of the proposed device. It will be suppressed if the range of the scattering potential is larger than the lattice constant but  smaller than the electron wavelength. This range also allows to model the scattering potential  as a $\delta$ function in the continuum limit \cite{AN98}. The best candidate for this would be adsorbed noble-gas atoms such as Ar or Xe, whose potential is intrinsically short-ranged. These atoms can be placed in the chosen positions for the experimental setup with the same STM tip (the process of single-atom manipulation with STM \cite{ES90} is by now a well controlled procedure \cite{TLH08}). Since these  impurities  have lower migration barriers than covalently bond impurities, the experiment will be better controlled at low temperatures to ensure that the impurities will not  move during the experiment \cite{WKL09}. Ionised impurities such as K with long-ranged Coulomb potentials can also be considered if a finite number of (gate-induced) carriers are added  to screen their effects. 

The validity of the semiclassical approximation requires an electron wavelength much smaller than the magnetic length. The electron wavelength in graphene is $\lambda = 2\pi v_F/E$, where $v_F \simeq 10^6 \; \rm m \cdot s^{-1}$ and $E$ is the energy measured from the Dirac point. We consider $E=0.4-2$ $eV$, with the largest $\lambda \simeq 10$ nm for $E = 0.4$ $eV$. (These energies are  routinely available in STM experiments \cite{LBetal10}) . At higher energies in the case of graphene trigonal warping comes into play but this deviation of the band structure from circular shape is not a problem for the setup and can be easily taken care of. 

The distortions described in Eq. \eqref{us}  produce an effective magnetic field that can be safely estimated to be $B_\text{eff} \sim \kappa (u_{00}/\sigma^2)\Phi_0$. This gives rise to an effective magnetic length $a_B \sim \sigma/(\kappa u_{00})^{1/2}$. For $\sigma=6$ nm and $u_{00}\lesssim$ 0.1 nm as considered before we have $a_B \gtrsim \lambda_{E=0.4eV}$, which makes it possible to work within the semiclassical approximation as we did. Note also that these parameters make it possible to reach fluxes $\phi \sim B_{eff}\sigma^2 $ such that $\phi \pi/\phi_0$ is of order 1. Similar numbers are obtained for the out-of-plane distortions described by Eq. \eqref{gaussian} with $A$= 0.5 nm and $\sigma$=6 nm as we considered before.

\section{A summary}

As a summary, in this work we put together two very beautiful physical facts: the Aharonov-Bohm effect, and the physical reality of the fictitious magnetic fields associated to elastic deformations or curvature in graphene. The maturity of the graphene field and the experimental capability to produce and manipulate large samples of very high quality together with the technical advances in the STM devices makes it possible to perform a new and very original AB experiment. As a byproduct and based on the physics described, a device can be made to measure micro strains in graphene samples at nanometer distances using an STM interferometer. The cases provided in the paper are only illustrative examples of the possible devices and applications of the setup.

\section{Acknowledgments}
\label{thank}
A.C. thanks J.M. G\' omez-Rodr\' iguez for inspiring discussions without which this work would not
have been initiated. We also thank R. Aguado, E. V. Castro, A. Geim, A. G. Grushin, F. Guinea and M. I. Katsnelson  for very positive and useful comments on the manuscript. We acknowledge support from
MEC (Spain) through grant FIS2005-05478-C02-01. F.J. acknowledges support from NSF through Grant No. DMR-1005035.

\section*{References}
\bibliography{AB}

\section{Supporting information}
\section{Supporting information}
\subsection{Fictitious gauge fields and the continuum model for graphene.}

As it is well known by now, the $\pi$-electrons of graphene that  are responsible for the electronic properties can be well described by a   tight binding model in the Honeycomb lattice \cite{W47,SW58}. The   Honeycomb lattice can be seen as two interpenetrating triangular   sub lattices. It has two points per unit cell. Due to this special   topology, the Bloch function has two components representing the   electronic probability at each sub lattice. The system at half filling   instead of a Fermi line has six Fermi points located at the corners of   the hexagonal Brillouin zone. Only two of them are inequivalent and can be chosen as $\mathbf K_1 =(4\pi/3\sqrt{3},0) $ and $\mathbf  K_2= -\mathbf K_1$. Expanding the dispersion relation around one of the Fermi points, say   $\mathbf K_1$ ($k_i=K_{1i}+\delta k_i$), one arrives to a continuum model for the low energy electronic   excitations  described by the equation  
\begin{eqnarray}
{\cal H}_1\sim -\frac{3}{2}ta \left(
\begin{array}{cc} 0 & \delta k_x+i\delta k_y
\\\delta k_x-i\delta k_y  & 0 \end{array} \right),
\label{contin}
\end{eqnarray}
where   $a\sim 1.42 $ {\AA} is the lattice constant and $t\sim 3$ $eV$ is the tight binding    From ${\cal H}/a$ in the limit ${a\to 0}$, $t.a$= const., one gets in the continuum  limit the massless relativistic Dirac Hamiltonian
\beq
H_1=\hbar v_F {\boldsymbol \sigma} \cdot {\mathbf k},
\label{dirach}
\eeq
where $\sigma_i$ are the Pauli matrices and $v_F = 3ta/2 \sim c/300$ is the   Fermi velocity of the electrons. The same expansion around the other Fermi point gives rise to a time   reversed Hamiltonian: ${H}_2=\hbar v_F(-\sigma_x k_x+\sigma_y k_y)$.   Notice that the two effective low energy Hamiltonians are related by   the change ${\boldsymbol \sigma}\leftrightarrow - {\boldsymbol \sigma^*}$.

The two Fermi points are referred to as  ``valleys" in a   semiconductor language. In the absence of interactions or disorder   mixing them, the valley quantum number is a degeneracy of the   spectrum. We must note that time reversal is not an invariance of the   effective hamiltonian around a given Fermi point but only of the full   system with the two Fermi points. The scattering potentials and effective gauge   fields invoked in this work may act on the valleys on a non-trivial   manner. In the case of the scattering potentials, since the two Fermi points are   far apart in momentum space ($|\mathbf K_2 -\mathbf K_1 |\sim 2K$), to avoid  intervalley scattering their range  has to be larger than the lattice unit $a$,
which is a reasonable assumption for the most common types of scatterers.

Concerning the effective gauge fields associated either to elastic   deformations or to intrinsic curvature of any origin, the crucial fact   is the they couple to each Fermi point with opposite signs. The   effective gauge fields can be obtained directly from the tight binding   model if  the derivation of the continuum model is done with arbitrary   nearest neighbour hoppings $t_i$ different in the three directions    \cite{SA02}. The Hamiltonian around the Fermi point $\mathbf K_1$ is
\begin{equation}
H_1=\hbar v_F {\boldsymbol \sigma} \cdot \left(-i{\nabla} - \mathbf { A}%
\right) ,  \label{df1}
\end{equation}
where $A_x =\frac{\sqrt{3}}2\left(t_3-t_2\right)$ , $A_y =\frac 12\left( t_2+t_3-2t_1\right)$. The effective potential has opposite sign around the second Fermi   point. The connection with elasticity theory    is obtained by assuming that the anisotropy in the    hoppings is due to atomic displacements of the positions ${\mathbf u}$ of   the atoms: $t_i=t+\delta t_i$, with $\delta t_i\sim u_i$. This defines the fictitious vector potential $\bf A$ given in Eq. \eqref{A}.

In the geometric approach to smooth curvature \cite{JCV07} the sign of the coupling of the effective vector fields depends on the definition of the spin connection and is determined by the   product of gamma matrices in it. As the effective Hamiltonians around   each of the Fermi points differ in the sign of only one of the gamma matrices  the spin connection has   opposite signs in the two valleys. In this approach the fictitious magnetic field is related directly to the  intrinsic curvature of the sample.

\subsection{STM interferences in the density of states}

We now show how the quasiparticle interference pattern produced by two impurities in the LDOS is modified in the presence of a fictitious gauge field in a way analogous to the AB effect. The derivation mainly proceeds along the steps of Ref. \cite{CP09}, though the fact that we have to deal with Dirac fermion Green's functions and the valley degree of freedom in graphene introduces additional subtleties that have to be worked out carefully. This is shown in detail in the following.  

We first revise the case of zero strain, for one valley. The LDOS measured by the STM tip at the point $\mathbf{r}$ for the energy $\omega$ from the Fermi level can be written as 
\beq
N({\bf r}, \omega)=-\frac{2}{\pi}{\rm Im}tr G({\bf r}, {\bf r};\omega).
\eeq
where $G$ is the electron Green's function in graphene. For a clean flat graphene sample one gets the (zeroth order) Green's function 
\beq\label{green}
G_{0}(\mathbf{r}_1,\mathbf{r}_2;\omega)  = -\frac{i\omega}{4v_F}\left[H_0(\omega|{\bf r_1}-{\bf r_2}|)+i\frac{{\bf \sigma}({\bf r_1}-{\bf r_2})}{|{\bf r_1}-{\bf r_2}|}H_1(\omega|{\bf r_1}-{\bf r_2}|)\right],
\eeq
where $H_0$ and $H_1$ are Hankel functions \cite{JCV07}.  The scatterers considered in the text can be associated with a scattering potential  of the form $U(\mathbf{r})= U_0\left[\delta(\mathbf{r}-\mathbf{r_1})+\delta(\mathbf{r}-\mathbf{r_2})\right] \mathcal I$, where $\mathcal{I}$ represents the identity in sub lattice and valley space. The Green's function can be expanded perturbatively in the scattering strength $U_0$ as 
\beq
\label{dosseries}
G({\bf r},{\bf r})= G_0({\bf r},{\bf r}) + \int dr' G_0({\bf r},{\bf r'})U({\bf r'})G_0({\bf
r'},{\bf r})+\cdots
\eeq
dropping the energy dependence since the scattering is elastic.  This gives rise to the term $N_{A=0}({\bf r},\omega)$ in Eq. \eqref{oscdos}, whose precise form is unimportant for our purpose. 

Consider now the case of non-zero strain. Our purpose is to show that the contribution to the LDOS due to the gauge field can be recast in the simple  interference-like term $N_{{\rm loop}} \left(\cos\Phi-1\right)$ that appears in Eq. \eqref{oscdos}. As mentioned in the text, $N_{{\rm loop}}$ is the part of $N_{A=0}$ generated by all diagrams associated with scattering paths in which the two impurities are involved (scattering loops). To see this, we compute the same perturbative series, but where the building blocks are now the Green's functions in the presence of the strain-induced gauge field. In the low-field regime, we can use a semiclassical approximation \eqref{G0A} for such a Green's function It is straightforward to show that terms involving one impurity only remain unchanged because these phase factors cancel. The loop terms are however modified. For example to second order we have
\beq\label{secor}
\delta G^{(2)}(\mathbf{r},\mathbf{r}) = U_0^2 tr \left[
G_{0}(\mathbf{r},\mathbf{r_1})G_{0}(\mathbf{r_1},\mathbf{r_2})G_{0}(\mathbf{r_2},\mathbf{r})e^{i
\frac{\pi
\Phi}{\phi_0}}+G_{0}(\mathbf{r},\mathbf{r_2})G_{0}(\mathbf{r_2},\mathbf{r_1})G_{0}(\mathbf{r_1},
\mathbf{r})e^{-i\frac{\pi \Phi}{\phi_0}}\right],
\eeq 
where $\Phi$, obtained from the line integral $\int_{\bf r}^{{\bf r}_1} + \int_{{\bf r}_1}^{{\bf r}_2} +\int_{{\bf r}_2}^{\bf r}$, is just the flux of the pseudo magnetic field through the triangle.  In the case of electrons in a normal metal, the Green's functions are scalars which moreover satisfy $G({\bf r_1},{\bf r_2})=G({\bf r_2},{\bf r_1})$, so a cosine factor is obtained from this expression directly. The absence of these two properties makes the calculation in our case more involved. 

We will first consider the second order case given by (\ref{secor}) and then proceed to arbitrary order. We want to show that the two terms in the expression (\ref{secor}) are actually equal (aside from the phase factors). For this, we first employ $G({\bf r_1},{\bf r_2}) = \sigma^3G({\bf r_2},{\bf r_1})\sigma^3$ and the cyclic property of the trace to show that the second term can be rewritten as
\beq
trG_{0}(\mathbf{r},\mathbf{r_1})G_{0}(\mathbf{r_2},\mathbf{r})G_{0}(\mathbf{r_1},\mathbf{r_2}).
\eeq 
We now commute the second and third Green's functions and obtain
\beq
\delta G^{(2)}(\mathbf{r},\mathbf{r}) = U_0^2 tr \left[
G_{0}(\mathbf{r},\mathbf{r_1})G_{0}(\mathbf{r_1},\mathbf{r_2})G_{0}(\mathbf{r_2},\mathbf{r})
2\cos\frac{\pi\Phi}{\phi_0}+
G_{0}(\mathbf{r},\mathbf{r_1})\left[G_{0}(\mathbf{r_2},\mathbf{r}),
G_{0}(\mathbf{r_1},\mathbf{r_2})\right]\right],
\eeq 
The commutator in this expression can be computed with (\ref{green})
\beq
\left[G_{0}(\mathbf{r_2},\mathbf{r}),G_{0}(\mathbf{r_1},
\mathbf{r_2})\right] = \frac{\omega^2}{4v_F}[\sigma^i,\sigma^j]\frac{(r_2-r)^i( r_1-r_2)^j}{|{\bf r_2}-{\bf r}||{\bf r_1}-{\bf r_2}|}H_1(\omega|{\bf r_2}-{\bf r}|)H_1(\omega|{\bf r_1}-{\bf r_2}|).
\eeq
Since $[\sigma^i,\sigma^j] = 2i \epsilon^{ij}\sigma^3$ for $i,j=1,2$, we have
\beq
trG_{0}(\mathbf{r},\mathbf{r_1})\left[G_{0}(\mathbf{r_2},\mathbf{r}),G_{0}(\mathbf{r_1},\mathbf{r_2})\right] \propto tr  G_{0}(\mathbf{r},\mathbf{r_1})\sigma^3 = 0 ,
\eeq
so this proves that at second order the loop contribution is 
\beq
\delta G^{(2)}(\mathbf{r},\mathbf{r}) = U_0^2 tr \left[
G_{0}(\mathbf{r},\mathbf{r_1})G_{0}(\mathbf{r_1},\mathbf{r_2})G_{0}(\mathbf{r_2},\mathbf{r})2\cos
\frac{\pi\Phi}{\phi_0}\right].
\eeq 
We now consider a general term in the loop series, paired with the corresponding term where the path is traversed in the opposite direction. A generic term of this sort will contain multiple bouncing off a single impurity as well as back-and-forth scattering between the two impurities. The first type of terms are easily re summed in the T-matrix of the impurity (the energy index is also omitted)
\beq
T = U_0 + U_0G({\bf r_n},{\bf r_n})U_0 + \cdots =  \frac{U_0}{1-U_0G({\bf
r_n},{\bf r_n})}.
\eeq
This expression is formally singular because of the real part of $G({\bf r_n},{\bf r_n})$. This has to be regularised, either by defining the real part in terms of the Kramers-Kronig transform of imaginary part, regularised with the bandwidth, or simply by an explicit computation in momentum space
\beq
G({\bf r_n},{\bf r_n}) = \int d^2 k \frac{\omega + {\bf \sigma}{\bf k}}{\omega^2-k^2 + i\epsilon} =
\omega \left( \log \frac{\omega^2}{\Lambda^2} + i\pi\right)
\eeq
With $\Lambda$ a momentum cutoff of the order of the lattice spacing. After this resummation, a generic term in the series will look like
\beq
\delta G^{(2m+2)}(\mathbf{r},\mathbf{r}) =tr
G_{0}(\mathbf{r},\mathbf{r_1})T_1G_{0}(\mathbf{r_1},\mathbf{r_2})T_2
\Big[G_{0}(\mathbf{r_2},\mathbf{r_1})T_1G_{0}(\mathbf{r_1},\mathbf{r_2})T_2\Big]^m
G_{0}(\mathbf{r_2},\mathbf{r})e^{i\frac{\pi\Phi}{\phi_0}}+
(1\leftrightarrow 2)e^{-i\frac{\pi\Phi}{\phi_0}}.
\eeq
 Applying the same trick as before, and noting that the T-matrix is proportional to the identity, this turns into
\begin{eqnarray*}
\delta G^{(2m+2)}(\mathbf{r},\mathbf{r}) =tr
G_{0}(\mathbf{r},\mathbf{r_1})T_1G_{0}(\mathbf{r_1},\mathbf{r_2})T_2
\Big[G_{0}(\mathbf{r_2},\mathbf{r_1})T_1G_{0}(\mathbf{r_1},\mathbf{r_2})T_2\Big]^m
G_{0}(\mathbf{r_2},\mathbf{r})e^{i
\frac{\pi
\Phi}{\phi_0}}+ \\
tr
G_{0}(\mathbf{r},\mathbf{r_1})T_1G_{0}(\mathbf{r_2},\mathbf{r})G_{0}(\mathbf{r_1},\mathbf{r_2})T_2
\Big[ G_{0}(\mathbf{r_2},\mathbf{r_1})T_2G_{0}(\mathbf{r_1},\mathbf{r_2})T_1\Big]^m e^{-i
\frac{\pi
\Phi}{\phi_0}}.
\end{eqnarray*}
So now we only need to commute the second Green's function to the right until it reaches the last place. As before, each step will produce a commutator proportional to $\sigma^3$. For example, the first will be 
\begin{eqnarray*}
tr
G_{0}(\mathbf{r},\mathbf{r_1})T_1\sigma^3 T_2
\Big[ G_{0}(\mathbf{r_2},\mathbf{r_1})T_2G_{0}(\mathbf{r_1},\mathbf{r_2})T_1\Big]^m e^{-i
\frac{\pi
\Phi}{\phi_0}}.
\end{eqnarray*}

Now the only non-zero traces in this expression contain $\sigma^3$ and an even number of $\sigma^i(r_a-r_b)^i$ with $i=1,2$ and $(r_a-r_b)$ the arguments of the Green's functions in the trace. All terms coming from the trace of such an expression will necessarily contain either vector products of proportional vectors $(r_1-r_2)^i(r_2-r_1)^j\epsilon^{ij}$, which vanish, or pairs of products of the type $(r-r_1)^i(r_1-r_2)^j\epsilon^{ij}+(r-r_1)^i(r_2-r_1)^j\epsilon^{ij}$, which also vanish. This can be shown to happen for all steps in the commutation, so that the two paths are indeed equivalent except for the phase factor. 

Finally, a resummation of the back-and-forth terms can also be done, which leads to the final expression
\beq
N({\bf r}, \omega)=N_{U=0}({\bf r}, \omega) + N_{{\rm loop}}({\bf r}, \omega)
\left(\cos\Phi-1\right),
\eeq
with
\beq
N_{{\rm loop}}({\bf r}, \omega) = \text{Im tr} 
G_{0}(\mathbf{r},\mathbf{r_1})W T_1G_{0}(\mathbf{r_1},\mathbf{r_2})T_2G_{0}(\mathbf{r_2},\mathbf{r})
\eeq
and
\beq
W = \frac{1}{1-T_1G_{0}(\mathbf{r_1},\mathbf{r_2})T_2G_{0}(\mathbf{r_2},\mathbf{r_1})}
\eeq

This calculation has been done  for one valley. The contribution from the second one is  obtained by changing  $\Phi \rightarrow -\Phi$ and we get  the same result because the cosine is even.

\end{document}